\documentclass[aip,jcp,a4paper, preprint]{revtex4-1}

\usepackage{placeins}
\usepackage{graphicx}
\usepackage{dcolumn}
\usepackage{bm}
\usepackage[]{placeins}
\usepackage{subfigure}
\usepackage{amsmath}

\newcommand{\LJs}{BLJM65~}
\newcommand{\LJl}{BLJM1560~}
\newcommand{\kB}{k_\mathrm{B}}
\newcommand{\Fc}[1]{{F_{\mathrm{c}, #1}}}
\newcommand{\av}[1]{{\langle #1 \rangle}}

\begin{document}
\title{Microrheology of supercooled liquids in terms of a continuous time random walk}
\date{\today}
\author{Carsten F. E. Schroer}
\email{c.schroer@uni-muenster.de}
\author{Andreas Heuer}
\email{andheuer@uni-muenster.de}
\affiliation{Westf\"alische Wilhelms-Universit\"at M\"unster, Institut f\"ur physikalische Chemie, Corrensstra\ss e 28/30, 48149 M\"unster, Germany}
\affiliation{NRW Graduate School of Chemistry, Wilhelm-Klemm-Stra\ss e 10, 48149 M\"unster, Germany}

\begin{abstract}
Molecular dynamics simulations of a glass-forming model system are performed under application of a microrheological perturbation on a tagged particle. The trajectory of that particle is studied in its underlying potential energy landscape. Discretization of the configuration space is achieved via a metabasin analysis. The linear and nonlinear responses of drift and diffusive behavior can be interpreted and analyzed in terms of a continuous time random walk. In this way the physical origin of linear and nonlinear response can be identified. Critical forces are determined and compared with predictions from literature.
\end{abstract}

\maketitle

\section{Introduction}

In microrheological experiments a physical system is brought out of equilibrium by a rather simple perturbation which is acting on a small fraction of its constituents~\cite{Cicuta2007}. These experiments are of specific interest because the application of a well-defined perturbation allows, first, to give predictions about the non-equilibrium behavior of the system and, second, to gain additional information about the equilibrium properties of the system by analyzing the transition to the non-equilibrium state.

In supercooled liquids in its (quasi-)equilibrium state one can observe a lot of unique dynamical properties like non-exponential relaxation or violation of the Stokes-Einstein relation~\cite{Binder2005}. Typically this is interpreted in terms of the presence of dynamic heterogeneities; see \cite{Heuer2008} and references therein. In recent years also the microrheological properties of supercooled liquids have been studied experimentally~\cite{Habdas2004,Wilson2009} as well as theoretically ~\cite{Reichhardt2006,Williams2006,Gazuz2009,Winter2012}. The occurrence of a nonlinear dynamical response was reported for different dynamical quantities. As expected for general grounds \cite{Williams2006}, the nonlinear response strongly increases at lower temperatures.

For the equilibrium dynamics of supercooled liquids it has been shown that a description of the dynamics in terms of transitions between metabasins (MB) yields important additional information about the nature of the slow dynamics at low temperatures \cite{Stillinger1995,DoliwaHopping,Souza2008}. It has been shown, that this discrete dynamics fulfills all criteria of a continuous time random walk (CTRW)~\cite{Berthier2005,Rubner2008} if applied to the description of a small system (O(10$^2$) particles). Most importantly, it turns out that the diffusivity of a small system only shows very small finite-size effects \cite{Rehwald2010} whereas the structural relaxation time displays large effects. This can be interpreted via facilitation effects \cite{Rehwald2010}. Going down in temperature may in principle modify the spatial and the temporal aspects of the CTRW. Interestingly, it turns out that the slowing down at low temperatures is exclusively determined by the temporal contributions \cite{DoliwaHopping}. Furthermore one can, e.g., express the complete wave vector dependence of the structural relaxation part of the incoherent scattering function in terms of the properties of the waiting time distribution, characterizing the CTRW \cite{Rubner2008}. Thus, the mapping of the MB dynamics on the CTRW is a powerful concept for the description of supercooled liquids. Unfortunately, the latter approach can be only used for small systems \cite{Heuer2008}. The qualitative reason is that very large systems can be decomposed into basically independent smaller systems \cite{Rehwald2010}. Thus, spatial information about the dynamic processes becomes essential which, however, is not available in configuration space. More formal arguments can be found in \cite{Heuer2008}.

 On a qualitative level the external force gives rise to a tilting of the potential energy landscape for the tagged particle. The key goal of the present work is to show how the CTRW properties are modified upon this tilting. This allows us to characterize the onset of non-linearity in the mobility of the tracer particles in terms of specific CTRW properties. Furthermore, it will be demonstrated that a small system, analogous to the equilibrium case, only displays small finite size effects, even in a high nonlinear regime. Therefore, a closer understanding of microrheological effects in small systems allows us to unravel the underlying physics of large systems as well.

 The paper is organized as follows: In Sect.~\ref{sim} we describe details of the simulation. Sect.~\ref{results} contains the results and their discussion. We conclude in Sect.~\ref{conclusion}.

\section{Simulations}
\label{sim}
We present the results of molecular dynamics (MD) simulations of a binary Lennard-Jones mixture, which consists of two types of particles, A and B, with an A:B ratio of $80:20$. This system is known to be a prototype of a glass-forming system~\cite{Kob1995}. To be able to simulate system sizes as small as 65 particles the cutoff radius is reduced to $1.8$ rather than $2.5$ (in dimensionless LJ-units)~\cite{DoliwaPEL}.
A microrheological perturbation is introduced to the system by randomly selecting one A type particle and pulling it with a constant force $F$ along a certain direction. To ensure that the system resides in a stationary state, the system is equilibrated under application of the force to the tracer particle. A constant temperature during our simulations is achieved by coupling the system to a Nos\'{e}-Hoover thermostat \cite{Nose1984}.

A key element of our analysis is the tracking of minima of the PEL which the system explores during its time evolution and which displays the impact of the external force. We access these minima by minimizing the particles coordinates in certain time intervals with respect to its potential energy ~\cite{DoliwaPEL}. Furthermore, following the procedure as outlined in Ref.~\cite{DoliwaPEL,Heuer2008} the minima are grouped together to metabasins (MB). As mentioned above the potential energy is minimized for the complete potential energy, containing also the effect of the external force. In practice it turns out that for somewhat larger forces ($F>10$) the minimization routine becomes instable. In any event, no larger forces are needed since the onset of non-linear effects occurs at much smaller values of $F$.

In MD simulations one generally prefers to simulate systems as large as possible to avoid any unphysical behavior due to the finite-size of the system. However, since the CTRW approach in configuration space is only valid for small system sizes \cite{Heuer2008} we focus on a small system with $N=65$ (BLJM65) which is close to the smallest system size without significant finite-size effects for the diffusivity\cite{Buechner1999}. Due to linear response theory this automatically implies that also the linear regime of the mobility in the microrheological setup should not display relevant finite-size effects. Whether or not this holds also for the non-linear regime will be checked by comparison with the simulations of a larger system with $1560$ particles and a geometry of $3:1:1$, the long side pointing along the force direction (BLJM1560).

\FloatBarrier
\section{Results and Discussion}
\label{results}
\subsection{Drift velocity in real space}
\label{driftReal}

One central quantity of the analysis of microrheological properties is the drift velocity $v$ the tracer particle obtains due to the interplay of permanent acceleration along the force direction and friction effects. Formally, the drift velocity is given by the stationary long-time limit of the displacement of the tracer particle parallel to the force direction $x_\parallel(t)$ relative to the time.

\begin{equation}
 v=\lim_{t\to\infty}\frac{x_\parallel(t)-x_{\parallel}(t_0)}{t-t_0}.
\end{equation}

\begin{figure}
 \includegraphics[width = 0.45\textwidth]{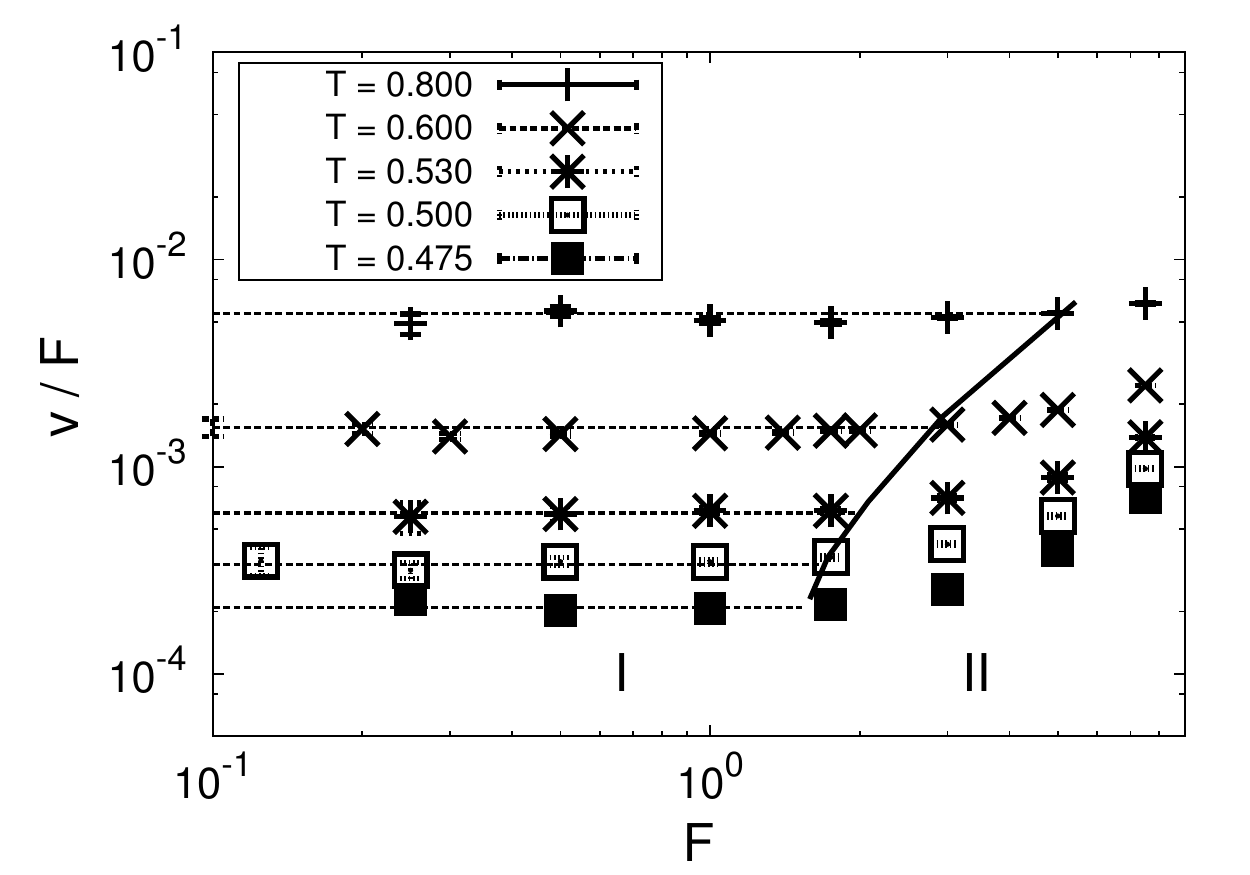}
 \caption{Drift velocity $v$ of \LJs. The dashed lines correspond to the expected linear response behavior (see Eq.~\ref{eqVLinRes}). The solid line indicates the critical forces $F_{c,v}$ at which the linear response regime (I) pass on to the nonlinear regime (II) (see Eq.~\ref{eqFc}).}
 \label{figDrift}
\end{figure}

For \LJs the force dependence of the drift velocity for different temperatures is shown in Fig.~\ref{figDrift}. In the range of forces of our investigation one can observe two regimes: a low force regime (I) in which the velocity increases linear with increasing force and a high force regime (II) which exhibit a nonlinear growth of velocity.

In the first regime, the velocity fulfills the linear response relation

\begin{equation}
\label{eqVLinRes}
 v=D_0\beta F
\end{equation}

(dashed lines in Fig.~\ref{figDrift}) in which $D_0$ stands for the one-dimensional diffusion constant in equilibrium and $\beta=\frac{1}{\kB T}$ for the inverse thermal energy. It is of particular interest, that the size of the linear regime is temperature dependent so that it decreases with decreasing temperature. Similar to the work of Williams et al.\cite{Williams2006}, we quantify this observation by fitting the curves with a symmetric power law

\begin{equation}
 \frac{v}{F}=a_2 F^2+a_0~.
\label{vPowLaw}
\end{equation}

Using this fit one can define a threshold force

\begin{equation}
\label{eqFc}
 \Fc{v} = \sqrt{\frac{0.1~a_0}{a_2}}
\end{equation}

by the criterion that the dynamical response differs more than $10\%$ from the expected linear response behavior. The values of these threshold forces are indicated as solid lines in Fig.~\ref{figDrift}. The critical forces are further discussed in Ref.\ref{critical}.

To analyze possible finite-size effects for the velocity we have compared the drift velocity of the \LJs and \LJl at a temperature of $T=0.475$. In the linear regime one observes that the velocity of the larger system slightly differs from the small system. According to linear response theory this just reflects analogous effects for the equilibrium diffusivity as reported in Ref.\cite{Rehwald2010}. Additional differences of the equilibrium diffusivities can result from hydrodynamic effects due to the different geometries. Most interestingly, after superimposing the velocity of the \LJs systems in the linear regime also the results for the nonlinear regimes are rather close. Especially the onset of nonlinear effects seems to be equal for both system sizes. Thus, understanding the nonlinearity for \LJs is sufficient to unravel the underlying physics also of large systems. In particular, this allows one to use the CTRW framework in configuration space for the elucidation of microrheological effects.

\begin{figure}
 \includegraphics[width = 0.45\textwidth]{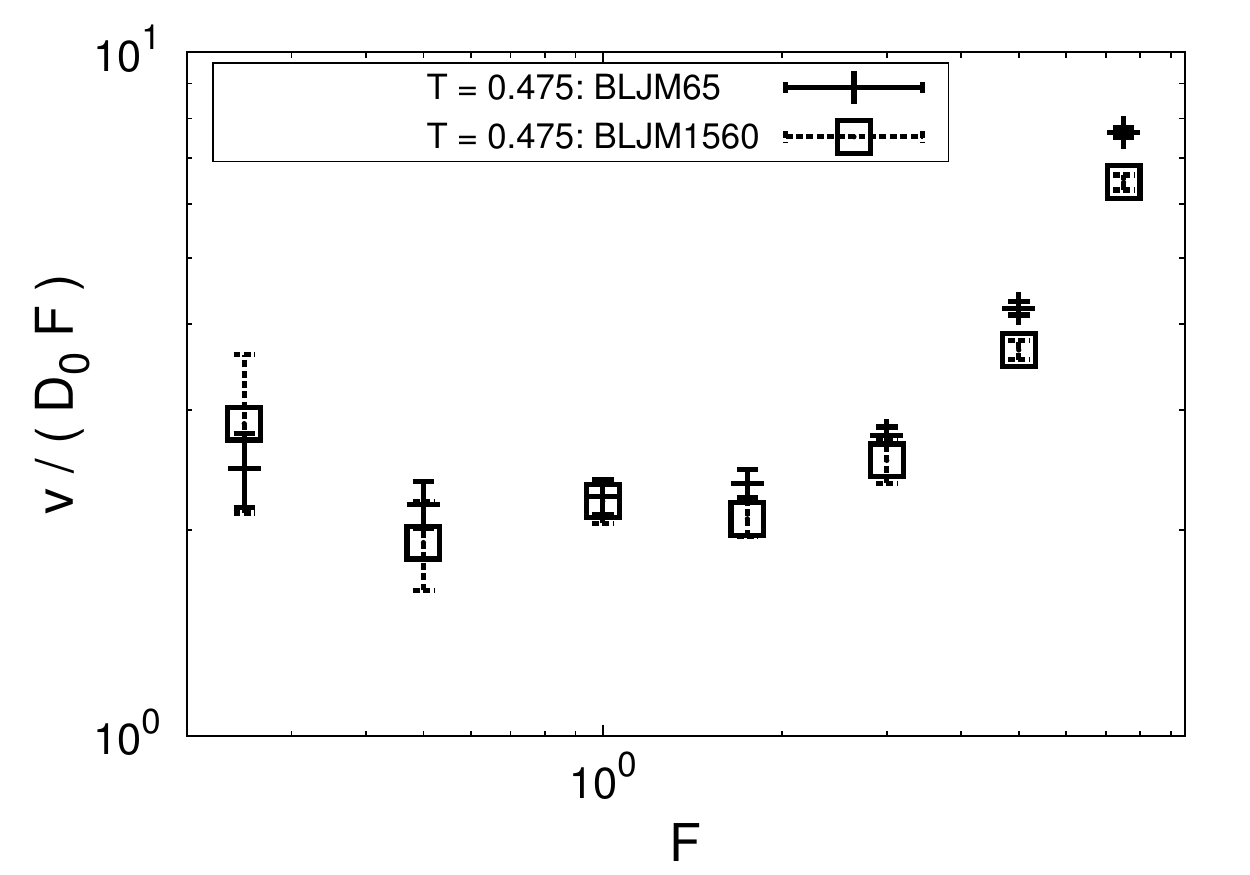}
 \caption{Drift velocity $v$ for \LJs and \LJl at a temperature $T=0.475$. $v$ was normalized via the particular equilibrium diffusion constants $D_0$.}
 \label{figFsize}
\end{figure}

\FloatBarrier
\subsection{Drift velocity in CTRW terms}
\label{driftCTRW}

In principle, any kind of discretization of the trajectories allows the decomposition of dynamical quantities in a spatial and a temporal part. In case of the drift velocity one can generally write

\begin{equation}
\label{eqDefv}
 v=\frac{\av{\Delta x_\parallel}}{\av{\tau}}.
\end{equation}

This relates the long time drift to the average displacement of the tracer particle along the force direction during a single elementary step, $\av{\Delta x_\parallel}$ and the average waiting time $\av{\tau}$. Because the waiting time cannot be affected by the direction of the applied force the value of $\av{\tau}$ can only depend on even powers of $F$. Therefore,  the linear response regime of small forces has to be related to the force dependence of the spatial part. However, in case of the nonlinear regime, it is a priori not clear, whether the spatial or temporal effects dominate the dynamical responses.

\begin{figure}
\includegraphics[width = 0.45\textwidth]{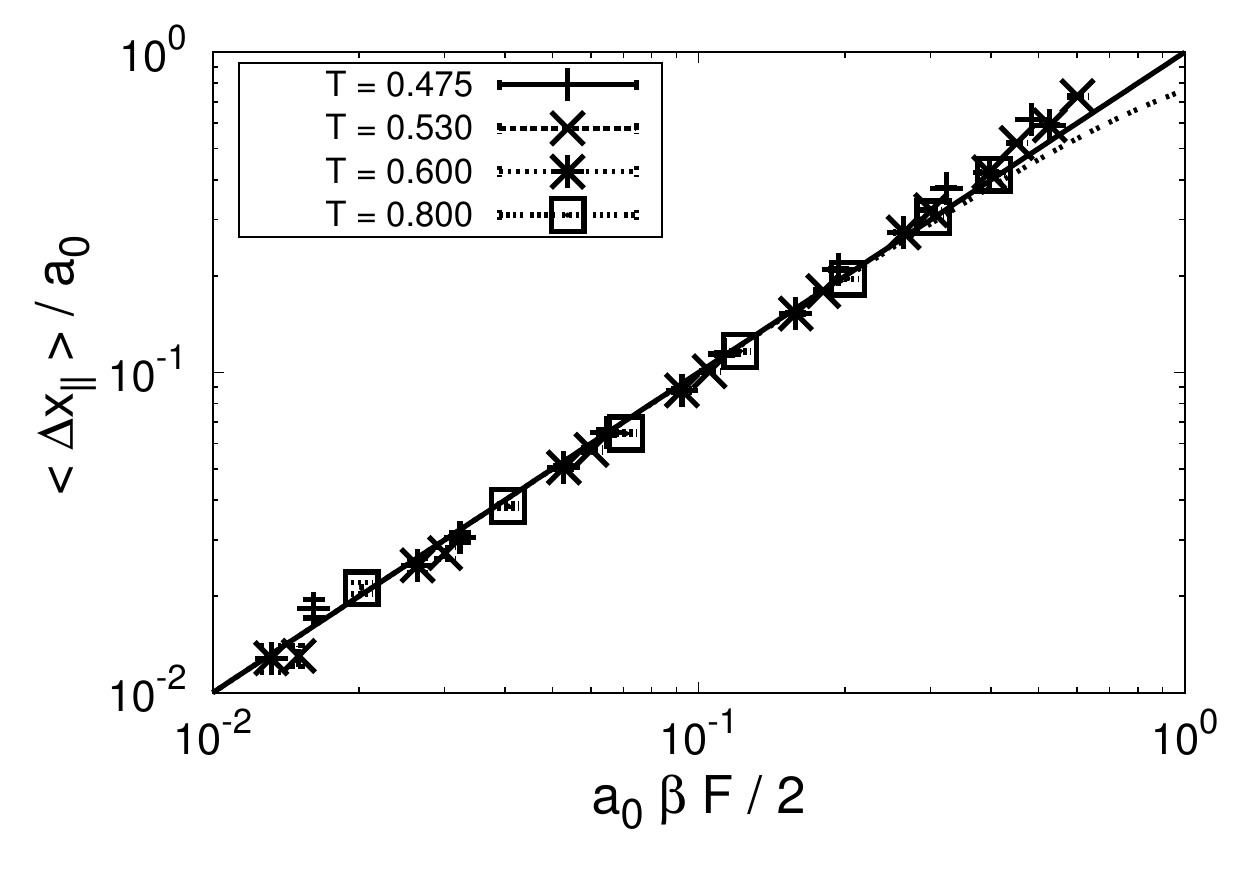}
 \caption{Average displacement of the tracer particle during one MB transition $\av{\Delta x_\parallel}$ as a function of the applied force $F$. The solid line indicate the linear response prediction by Eq.~\ref{eqDxLinRes} while the dashed line includes the theoretical prediction of nonlinear effects (Eq.~\ref{eqTanh})}
\label{figDxScaled}
\end{figure}

The force dependence $\av{\Delta x_\parallel}$ is shown in Fig.~\ref{figDxScaled}. One observes a linear scaling at low and intermediate forces. This behavior can be quantitatively understood by taking into account that the equilibrium diffusion constant $D_0$ can be expressed in CTRW terms \cite{DoliwaHopping} as

\begin{equation}
\label{eqDefD}
 D_0=\frac{a^2_0}{2\av{\tau}}~.
\end{equation}

The length scale $a_0$ refers to the one-dimensional averaged diffusive length a single A type particle moves during one MB transition in an equilibrium system. It is defined by the single particle displacement $x$ in one dimension after a large number of transitions $n$ via

\begin{equation}
\label{a2Def}
a^2_0 = \lim_{n\to\infty} \frac{\av{(x(n)-x(0))^2}}{n}.
 \end{equation}

As it was discussed in Ref.\cite{DoliwaHopping}, $a^2_0$ is a temperature independent quantity. Inserting Eq.~\ref{eqDefv} and Eq.~\ref{eqDefD} in Eq.~\ref{eqVLinRes} one obtains for the linear response regime

 \begin{equation}
 \label{eqDxLinRes}
\av{\Delta x_\parallel} = \frac{a^2_0}{2}\beta F~
 \end{equation}

which is indicated as a solid line in Fig.~\ref{figDxScaled}.

For the evolution of $\av{\Delta x_\parallel}$ in the nonlinear regime one can make a prediction by considering a one-dimensional periodic potential with a constant distance $a_0$ between two adjacent minima. Under application of the force, the jumping directions become biased so that one gets for the average displacement

\begin{equation}
\label{eqTanh}
\av{\Delta x_\parallel} = a_0\tanh{\frac{a_0}{2}\beta F}.
\end{equation}

 This particular ansatz was further discussed by Jack et al. \cite{Jack2008}. By linear expansion, Eq.~\ref{eqTanh} also includes the linear response relation (Eq.~\ref{eqDxLinRes}). However, from Eq.~\ref{eqTanh} one would expect a negative nonlinear response because the $F^3$ term of the expansion has a negative sign. In contrast, our numerical results display a slightly positive course. One can qualitatively understand this behavior by taking into account that in the theoretical ansatz $a_0$ is regarded as a force independent quantity. Especially at higher forces, one could expect that the diffusive length increases with increasing force as well (see also below). Unfortunately, one cannot immediately identify a specific length which displays the expected behavior to quantitatively reproduce the nonlinear growth of the particle displacement. Thus, we see that the spatial aspects of the hopping behavior at large forces cannot be described by this simple approach of equidistant minima.

In any event, the key conclusion from this analysis is the smallness of the nonlinear increase. Even at the lowest temperature and highest force it is smaller than a factor of $1.25$. As a conclusion, the spatial part of the drift velocity comes with a distinct linear and a very weak nonlinear response.

\begin{figure}
 \includegraphics[width = 0.45\textwidth]{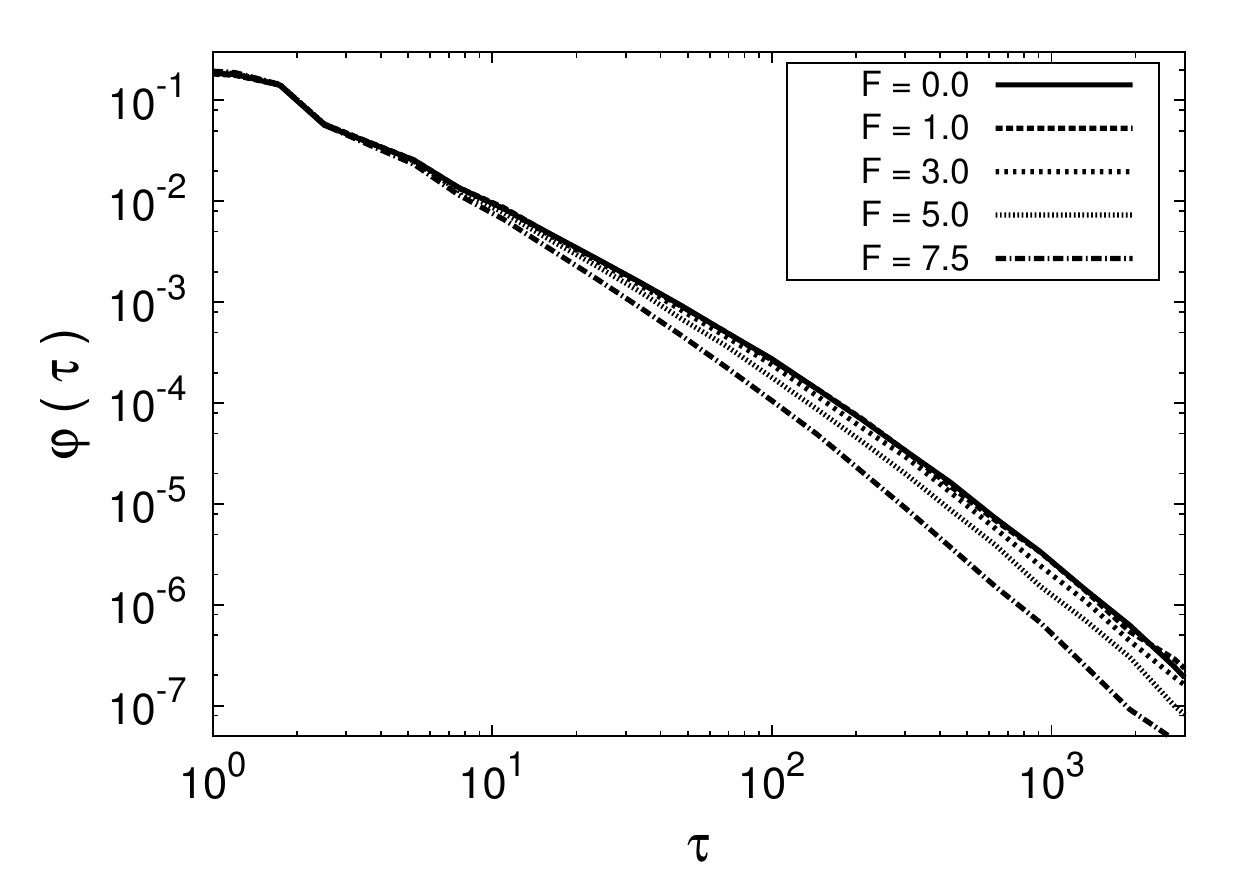}
 \caption{Distribution of MB waiting times $\tau$ during one MB transition at a temperature $T = 0.475$.}
 \label{figPhiTau}
\end{figure}

\begin{figure}
 \includegraphics[width = 0.45\textwidth]{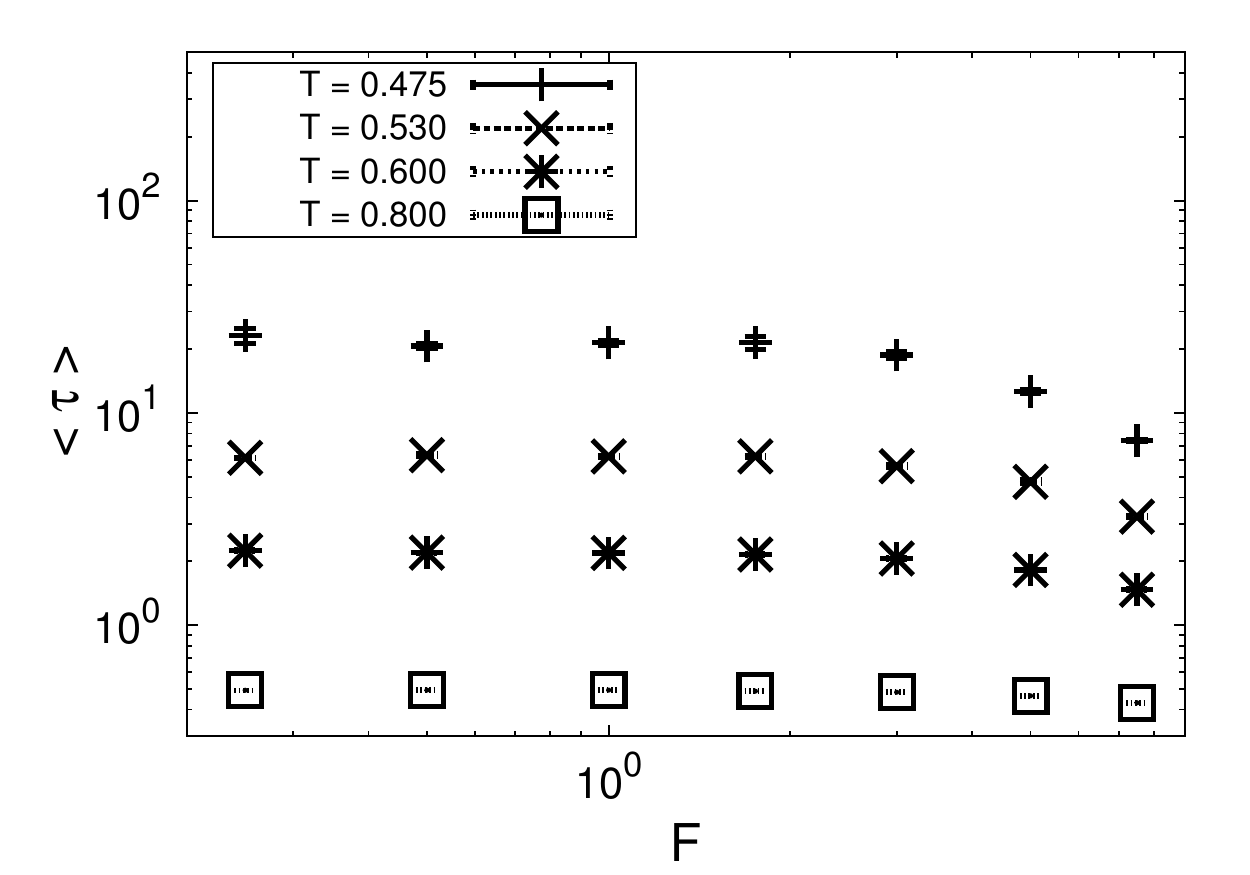}
 \caption{Average waiting times $\av{\tau}$ as a function of external forces $F$.}
 \label{figTauF}
\end{figure}

Concerning the temporal part we show the waiting time distribution $\varphi(\tau)$ in Fig.~\ref{figPhiTau}. In the case of small forces, the distribution of waiting times does not change so that the curves collapse as expected for the linear response regime. Indeed, in case of higher forces, long waiting times do not occur any more so that the distribution functions decay faster. This has major consequences for the average waiting time $\av{\tau}$ (see Fig.~\ref{figTauF}): While for small forces $\av{\tau}$ stays constant, one can observe a drop for forces higher than a critical force. We characterize the critical behavior of the temporal part by determining the critical force of the inverse average waiting time $1/\av{\tau}$ which is the relevant contribution to the drift velocity (see Eq.~\ref{eqDefv}). It turns out, that the critical force $\Fc{\frac{1}{\av{\tau}}}$ is identical to $\Fc{v}$. This fact underlines the major influence of the temporal part to the nonlinear behavior of the drift velocity. For further discussion in \ref{critical} we will thus not distinguish between these critical forces and restrict our analysis on $\Fc{\frac{1}{\tau}}$.

The transition to the nonlinear regime of $v$ can be regarded as a direct consequence of the nonlinear behavior of $\av{\tau}$. The observations of the different force dependencies of the spatial and the temporal part interestingly illustrate that the discretization of the trajectory in the MB approach also leads to a decomposition of linear and nonlinear contributions to the drift velocity: In case of small forces, $\av{\tau}$ stays constant so that the linear response can be completely related to the spatial part. The nonlinear regime, however, is governed by a decay of the average waiting time.

On a qualitative level this picture agrees with the idea of linear response theory as applied to a simple 1D cosine-potential. Naturally, $\av{\Delta x_\parallel}$ is proportional to the force in the linear regime. Thus, any variation of the waiting times, e.g. to a modification of the barriers, gives rise to higher-order effects in the resulting mobility. Stated differently, the nonlinearity mainly reflects the modification of the underlying PEL upon application of a large external force.

%\FloatBarrier
\subsection{Diffusion in CTRW terms}
\label{diffCTRW}

As it was already mentioned above (see Eq.~\ref{eqDefD}), the one-dimensional diffusion constant in equilibrium of a single particle is given in CTRW terms as the ratio of the isotropic diffusive length $a^2_0$ and the average waiting time $\av{\tau}$. Going towards driven system one has to distinguish between the diffusive processes parallel and perpendicular to the force direction which were characterized by the diffusion constants $D_\parallel$ and $D_\perp$, respectively. Because $\av{\tau}$ is a universal quantity, expected differences between parallel and perpendicular diffusion are related to the anisotropy of the respective length scales. In analogy to Eq.~\ref{a2Def}, we define the length scale in perpendicular direction $a^2_\perp$ as

%\FloatBarrier

\begin{figure}
 \includegraphics[width = 0.45\textwidth]{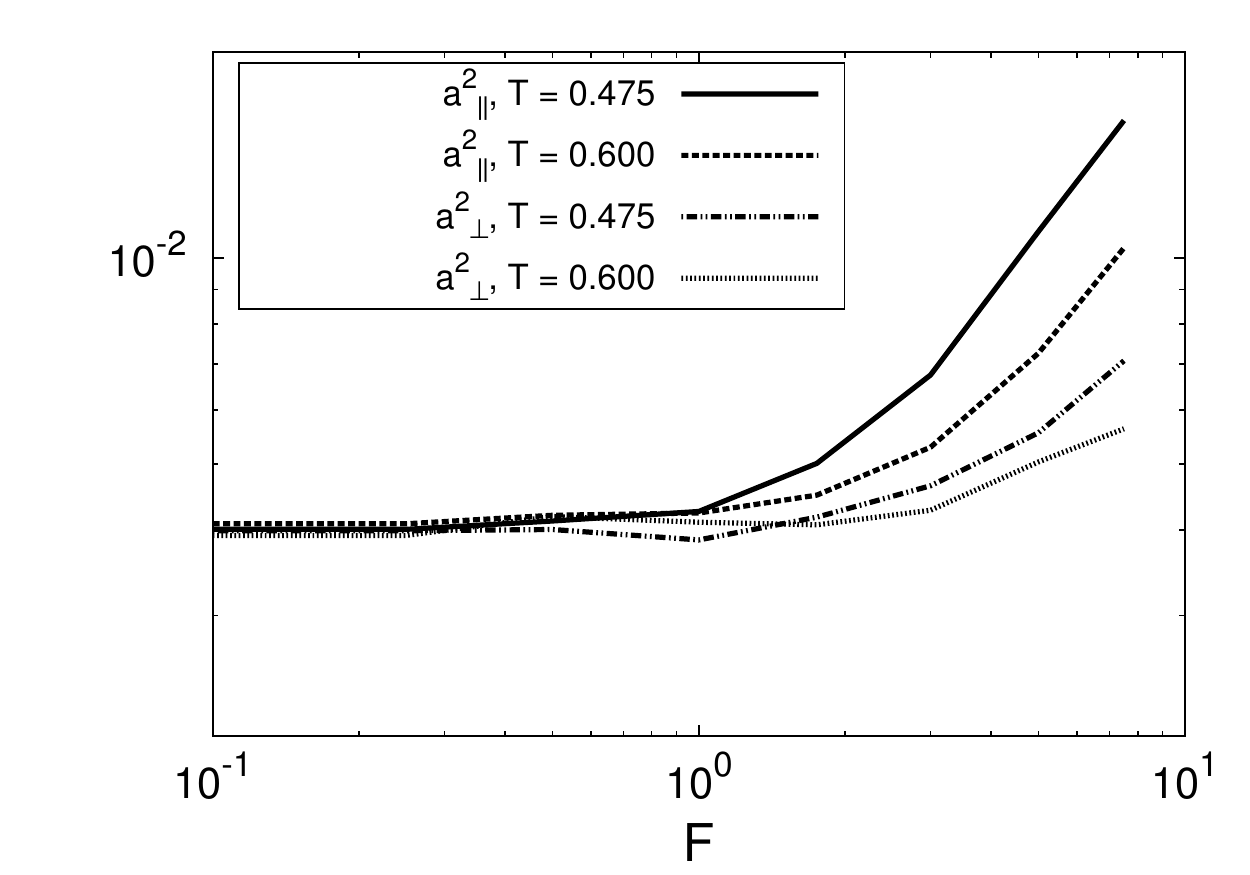}
 \caption{Diffusive length scales $a^2_\parallel$ (two upper lines) and $a^2_\perp$ (two lower lines) of the tracer particle parallel and perpendicular to the force direction as a function of the external force~$F$.}
 \label{figDiffLength}
\end{figure}

\begin{equation}
 a^2_\perp = \lim_{n\to\infty} \frac{\av{(x_\perp(n)-x_\perp(0))^2}}{n}
\end{equation}

and the parallel length scale $a^2_\parallel$ as

\begin{equation}
\label{eqDefa2parallel}
 a^2_\parallel = \lim_{n\to\infty} \frac{\av{(x_\parallel(n)-x_\parallel(0))^2}-\av{(x_\parallel(n)-x_\parallel(0))}^2}{n}.
\end{equation}

In the latter definition we subtracted the systematic shift of the tracer particle along the force direction. The corresponding length scales are shown in Fig.~\ref{figDiffLength}. One can observe that both, $a^2_\perp$ and $a^2_\parallel$, exhibit an increase with increasing force which is stronger for the parallel direction. Interestingly, although $a^2_\perp$ and $a^2_\parallel$ are temperature independent in equilibrium, corresponding to the linear regime, the degree of nonlinearity strongly depends on temperature. As already indicated above one may expect that the PEL properties change in the nonlinear regime. Since these modifications are expected to be strongly anisotropic it is not surprising that observables, defined parallel or orthogonal to the applied force, may behave differently.

\begin{figure}
\includegraphics[width = 0.45\textwidth]{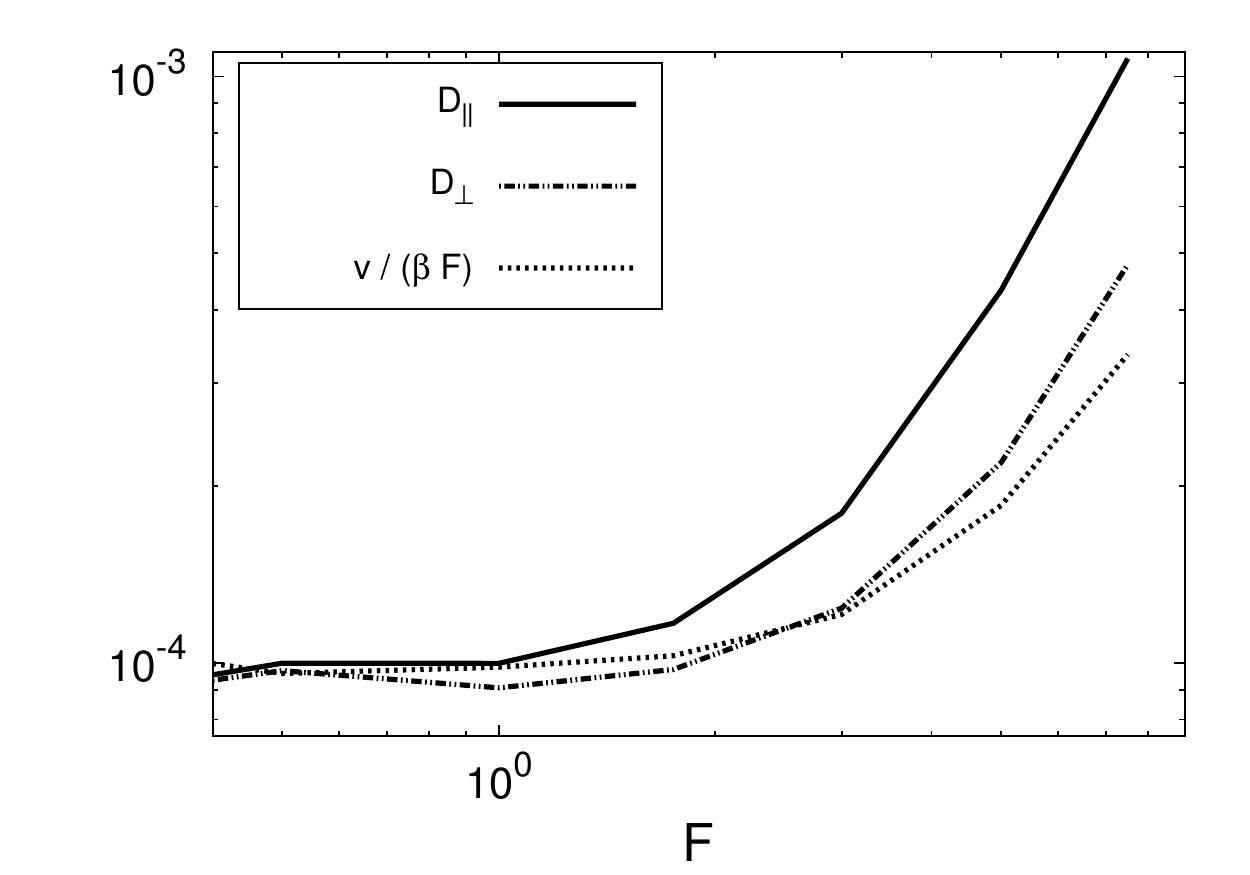}
 \caption{Nonlinear responses of the diffusion coefficients parallel ($D_\parallel$) and perpendicular ($D_\perp$) to the force direction and of the drift velocity $v$ at a temperature $T = 0.475$.}
 \label{figDiff}
\end{figure}

Dividing $a^2_\perp$ and $a^2_\parallel$ by $2\langle \tau \rangle$ one obtains the parallel and the orthogonal diffusion constants $D_\parallel$ and $D_\perp$, respectively. Since the denominator is always the same the different increase of $a^2_\parallel$ as compared to $a^2_\perp$ directly translates into the corresponding relative increase of $D_\parallel$ as compared to $D_\perp$. Furthermore, the weak force-dependence of $\av{\Delta x_\parallel}$ directly translates into a weaker force dependence of the mobility as compared to both diffusion constants. Actually, the observed relation $D_\parallel\gg D_\perp > \frac{v}{\beta F}$ (see Fig.~\ref{figDiff}) was already reported for the lithium dynamics in a lithium silicate system~\cite{Kunow2006}. This suggests that the force dependent anisotropy of the relevant length scales is a general property of disordered systems.

\FloatBarrier
\subsection{Critical forces}
\label{critical}

Finally, it is of interest to compare the nonlinear response of the different observables on a more quantitative level. Since we are mainly interested in the onset of nonlinearity an appropriate measure is the critical force as introduced in Eq.~\ref{eqFc}. Here we concentrate on $a^2_\parallel$, $a^2_\perp$ and $1/ \langle \tau \rangle$ as the key observables of the CTRW approach. The remaining variable $\av{\Delta x_\parallel}$ is omitted because its nonlinearity is extremely small. Furthermore, the critical force of the drift velocity $\Fc{v}$ is not discussed separately because, as it was mentioned above, it has been found to be identical with $\Fc{\frac{1}{\tau}}$.

\begin{figure}
\includegraphics[width = 0.425\textwidth]{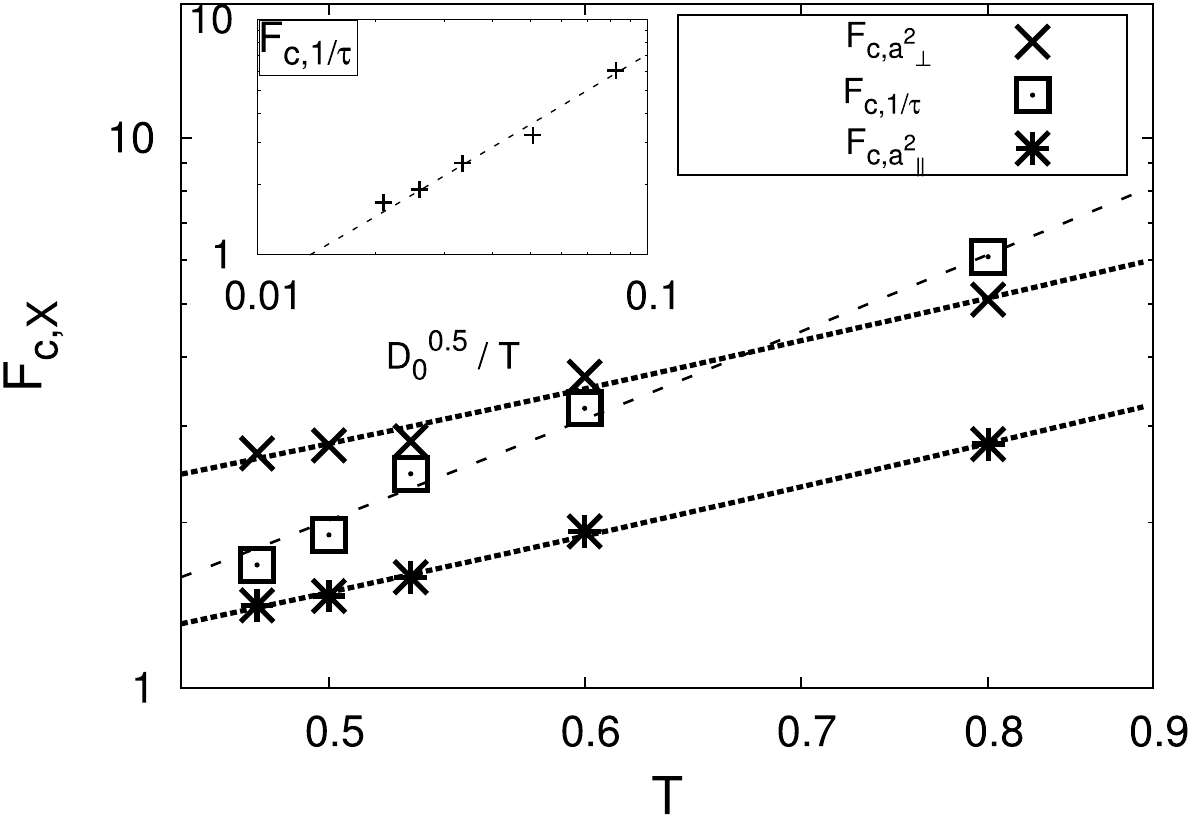}
  \caption{Critical forces $\Fc{X}$ of different dynamical quantities $X =$ $a^2_\parallel$, $a^2_\perp$ and $\frac{1}{\tau}$. The dashed lines indicate a behavior $\Fc{X}\propto T^{1.5}$ or rather $\Fc{X}\propto T^{2.5}$. Inset: $\Fc{\frac{1}{\tau}}$ as a function of $\frac{\sqrt{D_0}}{T}$ with the equilibrium diffusion constant $D_0$. The dashed line indicates a linear behavior.}
 \label{figFcAll}
\end{figure}

In Fig.~\ref{figFcAll} all critical forces are shown as a function of temperature. It can be seen that in general for a given temperature each observable has a different critical force. Furthermore the temperature dependence varies quite significantly. Whereas for the length scales the critical forces scale similarly to $\Fc{X}\propto T^{1.5}$, the critical force of the inverse waiting time roughly scales like $\Fc{\frac{1}{\tau}}\propto T^{2.5}$.

A theoretical prediction about the onset of nonlinearity can be found in \cite{Evans2005}. It was suggested by the authors that linear response relation of the occurrent flux in a driven system holds exactly when the steady state fluctuation theorem

\begin{equation}
 \frac{p(v(t)=A)}{p(v(t)=-A)}\propto exp(A\beta F t)
\end{equation}

is fulfilled. From this condition the the scaling

\begin{equation}
\label{eqEvans}
 \Fc{v}\propto\frac{\sqrt{D_0}}{T}
\end{equation}

 has been derived in which $D_0$ stands for the equilibrium diffusion coefficient. In the inset of Fig.~\ref{figFcAll} one can observe that $\Fc{\frac{1}{\tau}}$, and therefore naturally $\Fc{v}$ as well, fulfills the scaling prediction.

\section{Conclusion}
\label{conclusion} In the present paper we have analyzed the
dynamics of a driven single  particle in a supercooled liquid in
terms of a CTRW. In case of the drift velocity the discretization
of the MB approach allows one to identify the temporal part of the
CTRW as the key source of nonlinear response, whereas the spatial
part only exhibits a weak positive contribution. This positive
contribution can be qualitatively understood by the fact that the
distances between adjacent MB increase due to the application of
higher forces. Since the finite-size effects are small these
results also reflect the origin of the nonlinear response in the
much larger system.

In terms of a CTRW analysis we have  been able to determine the
relevant diffusive lengths of the tracer particle diffusion along
the directions parallel and perpendicular to the force. Among
others, this enables the definition of a parallel diffusion
constant which ís not directly accessible from real space
trajectories \cite{Winter2012}.

The force dependencies of parallel  diffusion, perpendicular
diffusion and drift velocity are significantly different. Thus,
there is no general critical force which designates a mutual
transition to a nonlinear regime. However, in terms of dynamical
quantities, one can regard the critical force of the inverse
waiting time as a key entity because of its relation to the
velocity (neglecting the minor nonlinear effects of the spatial
part). For this quantity one also observes the validity of the
theoretical scaling prediction by Evans et al..

This observation is of major conceptual  interest because the
theoretical prediction is based on steady state fluctuation
arguments while the CTRW waiting time is determined by the
underlying PEL, i.e. by the distribution of minimum energies and
saddle heights as well as the topography. Because the waiting time
decreases under the application of the external force, one can
assume that the force dependence of at least one of these
distributions is responsible for this behavior. It remains to be
shown which specific variations of the PEL give to the transition
for linear to nonlinear dynamics.

\section*{Acknowledgements}
This work was supported by DFG Research Unit 1394 "Nonlinear Response to Probe Vitrification". Furthermore, C. F. E. Schroer thanks the NRW Graduate School of Chemistry for funding and C. Rehwald and O. Rubner for many helpful discussions about this work.

%\bibliography{p1.bib}

%Merlin.mbs v4.21 2009-07-09.
%

\end{document}